\documentclass[prl, twocolumn, footinbib,a4paper, showpacs]{revtex4}
\usepackage[latin1]{inputenc}
\usepackage{amssymb}
\usepackage{amsmath}
\usepackage{amsbsy}
\usepackage{bm}
\usepackage{graphicx}

\begin{document}
\title{Take-off of small Leidenfrost droplets }

\author{Franck Celestini $^1$, Thomas Frisch$^2$ , Yves Pomeau$^3$ }

\affiliation{$^1$ Laboratoire de Physique de la Mati\`ere Condens\'ee, CNRS UMR 7366, Universit\'e de Nice Sophia-Antipolis,
   Parc Valrose  06108 Nice Cedex 2, France}
   
\affiliation{$^2$ Institut  Non Lin\'eaire de Nice, CNRS UMR 7735,  Universit\'e de Nice Sophia-Antipolis,
   1361 Routes des lucioles, Sophia Antipolis F-06560 Valbonne France}

\affiliation{$^3$  University of Arizona, Department of Mathematics, Tucson, AZ 85721 USA}
\begin{abstract}

We put in evidence the unexpected behaviour of Leidenfrost droplets at the later stage of their evaporation.  We predict and observe that, below a critical size $R_l$, the droplets spontaneously take-off due to the breakdown of the lubrication regime. We establish the theoretical relation between the droplet radius and its elevation. We predict that the vapour layer thickness increases when the droplets become smaller.
A satisfactory agreement is found  between the model and the experimental
results performed on droplets of water and of ethanol.

\end{abstract}

\pacs{ 47.55D-,68.03.-g}

\maketitle

The name of J.G. Leidenfrost (1715-1794) is still remembered because he was the first to publish the observation that a puddle of water dropped on a very hot surface divides into droplets, which stand each above the surface and slowly evaporate. At the end of the process, he noticed \cite{leiden} :
``... before the whole drop disappears. Which at last exceedingly diminished so that it can hardly any more be seen, with an audible crack, which with the ears one easily hears, it finishes its existence, and in the spoon [making the hot surface] it leaves a small particle of earth ..."  

Over the years, the Leidenfrost phenomenon has attracted the attention of many investigators, see for instance \cite{quere,linke,toro,lagubeau,eggers,gold,wurger}, but seemingly little interest has been paid to the observation by Leidenfrost of the final stage of the droplet evaporation, in particular what happens when the droplets become very small before finally disappearing.
 We investigate in this letter theoretically and experimentally the final regime of the Leidenfrost droplets.
  We show that when the droplets become smaller than a well defined radius, they  suddenly take-off due to the breakdown of the classical  Leidenfrost lubrication regime. They thus reach an elevation which is  much higher than their radius. 

\begin{figure}[!h]
  \includegraphics[width=.45\textwidth]{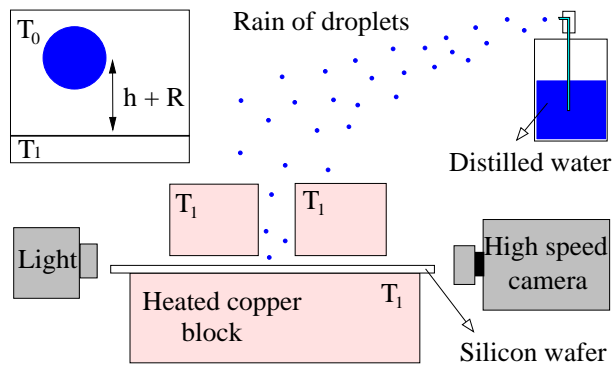}
  \caption{Experimental set-up. A cloud of sub-millimetric droplets is rained toward a silicon wafer inserted in a copper bloc kept at a controlled temperature $T_1$. The droplet of radius $R$ is at temperature $T_0$ and at a distance $h$ of the substrate. }
  \label{fig_set-up}
\end{figure}

Let us consider first an almost spherical drop of volatile liquid with radius $R$ and mass $M$ standing slightly above a hot flat surface with a vapour film of  thickness $h$ in between. Let $l$ be the horizontal width of this film, this length will be related to the other physical parameters of the problem. 

Our estimate will rely on the fact that, in a range of parameters, the horizontal length scale $l$ is much larger than $h$. Both $l$ and $h$ are less than $R$, the radius of the droplet. Furthermore, we shall consider a range of parameters where the vapour pressure in the film is negligible compared to Laplace's capillary pressure. Therefore the droplet remains almost exactly spherical in this regime. This will yield the condition (\ref{eq:Poiseuille.2}) derived below, valid under the constraint $h \ll R$.  This condition happens to be impossible to satisfy if the droplet becomes too small, that is if its weight is too small to balance the upward pressure in the gap between the hot plate and the droplet. Therefore, it is natural to guess that, when this happens, the droplet lifts off the plate. This defines a second regime where $h \gg R$. The scaling laws for this second regime are given in this letter and we show that the droplet reaches an higher elevation as it gets smaller and smaller by evaporating.

Energy conservation during the evaporation process of the droplet (namely Stefan's boundary condition on the liquid/vapour interface), which we assume to be just below the boiling temperature, and Fourier law for the heat transfer in the film yield the order of magnitude for the vertical velocity  of vapour $w$ near the surface of the droplet:
\begin{equation}
w= \frac{\lambda \delta T}{ h L \rho_v}
\mathrm{.}
\label{eq:Poiseuille}
\end{equation}
Here $\delta T= T_1-T_0$ is the temperature difference between the droplet and the hot plate, $\rho_v$ is the mass density of the vapour, $\lambda$ the heat conductivity in the vapour  and $L$ the latent heat of evaporation per unit mass.
 Lubrication theory, valid if  $l\gg h$, yields the magnitude of the horizontal velocity $u$ in term of the vertical velocity : $
u= \frac{l w}{h} $, where $l$ is the horizontal extent of the vapour film. From Poiseuille relation such a flow is driven by a horizontal pressure gradient scaling as
$ \delta P/l \sim \eta u/ h^2$. Therefore the typical gradient is of order 
$\delta P/l \sim w l \eta /h^3$. Using  now equation (\ref{eq:Poiseuille}) we find (replacing sign $\sim$ by $=$ for readability) : 
\begin{equation}
 \delta P= \eta  l^2 \frac{\lambda \delta T}{ h^4 L \rho_v}
 \mathrm{,}
\label{eq:Poiseuille.1}
\end{equation}

The liquid  drop is at mechanical equilibrium when $M g= \delta P l^2 $ namely when its weight is equal to the vertical component of the force due to viscous pressure in the gap between the droplet and the hot plate. Note that the contribution of the viscous stress to this vertical force is of the same order of magnitude as the one of hydrostatic pressure. Therefore we find using the mechanical equilibrium relation given above that :

\begin{equation}
 \frac{l^4}  {h^4}  = \frac{L \rho_v  M g}{\eta  \delta T \lambda}
 \mathrm{.}
\label{eq:Poiseuille.2}
\end{equation}

As stressed above the lubrication approximation  becomes invalid  when $h\sim l $. From the $M$ dependence of the right-hand side of equation (\ref{eq:Poiseuille.2}) we find that the lubrication approximation becomes invalid if the radius of the droplet is less than a critical drop radius $R_l$ defined by the condition $l = h$. The breakdown of the lubrication approximation will then occur for :
\begin{equation}
R_l = \left(\frac{  \eta  \delta T \lambda}{g L \rho_v \rho_l}\right)^{1/3} 
 \mathrm{.}
\label{eq:Poiseuille.3}
\end{equation}

\begin{figure}[!h]
  \includegraphics[width=.45\textwidth]{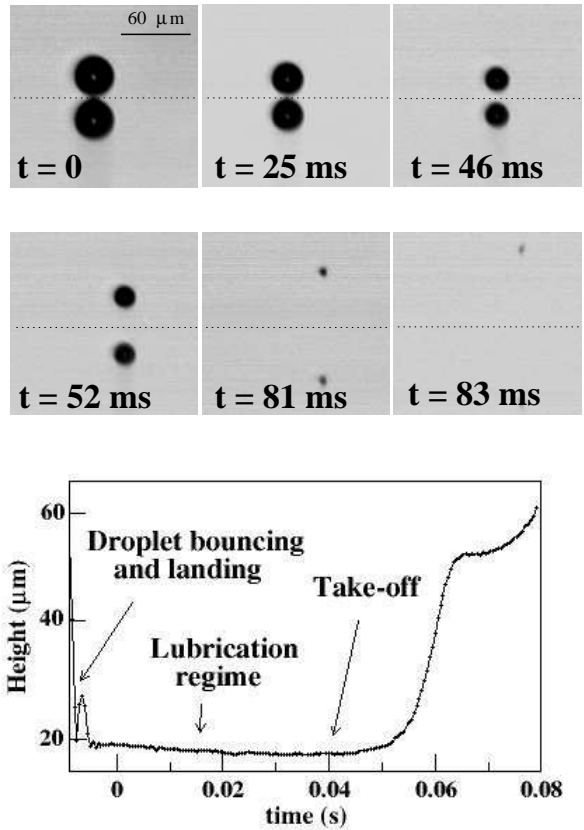}
  \caption{Top : Pictures of a Leidenfrost droplet and its reflected image on the silicon wafer.  Bottom : Height of the droplet as a function of time. After being dropped, it first bounces and lands on the substrate. The droplet then stands on the substrate in the lubrication  regime. It finally takes-off from the substrate. }
  \label{fig_ilus}
\end{figure}

 Therefore one expects that, as $R$ gets smaller than $R_l$, the droplet takes off from the hot plate. The range of values of $h \sim R \sim R_l$ is beyond reach of simple order of magnitude estimates and its analysis requires a solution of the full set of equations for the multi-phase Stokes flow and the heat transfer model. Let us consider the asymptotic regime where  $R$ is much less than $R_l$, where one expects the droplet to lift well above the hot plate ($h\gg R$). 
 In this regime, the equilibrium elevation of the droplet is reached when the gravity force is balanced by the dipolar component of the velocity field generated by the evaporation from the droplet surface and by its image with respect to the horizontal plane. This dipolar approximation is correct for $h\gg R$. Let $T_1$ the temperature at infinity and $T_0$ the  temperature of the droplet. The temperature field satisfying Laplace's equation around an isolated droplet reads : $T(r) = T_1 +(T_0-T_1) R/r$. The origin of the spherical coordinates ($r=0$) is taken at the center of the droplet. Let us consider now the temperature due to the image with the temperature field of order  $T_{im} \sim (T_1-T_0)R/2h$ near the physical droplet, with a nearly uniform vertical gradient. As justified below, in the limit of large $h$ the net vertical heat flux is of order of the derivative of $T_{im}$ with respect to $h$, namely $(T_1-T_0)R/h^2$. Using Stefan's law we find the order of magnitude of the net vertical velocity in the vapour near the droplet reads :
 
 \begin{equation}
  \frac{\lambda}{ L \rho_v}\frac{(T_1-T_0)R}{h^2}
  \mathrm{.}
 \end{equation}
 
   This relation is similar to equation (\ref{eq:Poiseuille}), but with $\delta T/h$ replaced by $(T_1-T_0)R/h^2$. Multiplying this velocity  by $\eta R$, as in Stokes drag law, we obtain the vertical force on the droplet.
 Balancing this force with gravity we find that, in the limit $h\gg R$ (namely for $R \ll R_l$), the dimensionless elevation of the droplet at equilibrium $h^*= h/R$ should scale as :
  
\begin{equation}
h^* =  (R_l/R)^{3/2} 
\mathrm{.}
\label{grandh}
\end{equation}

Before going further we have to analyse the constraint on the pressure in the gap: it must be negligible
compared to the Laplace's pressure in the droplet to ensure that it remains close to its spherical shape everywhere.
It can be shown \cite{note} that $l\sim \sqrt{Rh}$ is a fair estimate of the horizontal extent of the layer for which the viscous pressure contributes significantly to the vertical force.
Using eq.(\ref{eq:Poiseuille.1}) and this  relation  between $l$, $R$ and $h$, the pressure induced by the
Poiseuille flow in the gap is found to be
 of order $\delta p = g \rho_l \frac{R R_l^3}{h^3}$. It crosses the order of magnitude of Laplace's pressure inside the droplet when $\delta P = \frac{\sigma}{R}$. This happens when the radius becomes of order
$R_i = \left(\frac{\sigma}{g \rho_l}\right)^{2/7} R_l^{3/7}$.
 This relation is obtained using the above geometrical relation for $l$ and equation (\ref{eq:Poiseuille.2}) reformulated as $h/l = (R_l/R)^{3/4}$.
 This assumes that the capillary length $\left(\frac{\sigma}{g \rho_l}\right)^{1/2}$ is much larger than $R_l$, since otherwise the cross over would happen in a range of values of $R$ smaller than $R_l$: this is impossible because $R_i$ was derived under the assumption that the lubrication approximation holds in the gap between the sphere and the hot plate. The condition $R_i \gg R_l$ is fulfilled in the experiments described below, but there could be other situations at stakes for which this condition may not be fulfilled.   This gives us also the opportunity to address another feature of our experiments, namely the difference of behaviour between droplets of water and of ethanol. Since the quantity $R_i/R_l$ is independent on the radius of the droplet, but depends on the properties of the liquid and its vapour, the behaviour of Leidenfrost droplets is not unique up to rescalings, because it depends on a dimensionless ratio with different value depending on the liquid used. Let us also notice that for radii larger than $R_i$, the shape of the droplet becomes rather complex, because it depends on the (large) deformation of the droplet surface due to the Poiseuille pressure inside the vapour gap \cite{nagel,celestini}. This happens in a range of parameters outside of this study. 
  Therefore we focus on the range of parameters such that $R_l \ll R \ll R_i$. In this range, using equation (\ref{eq:Poiseuille.2}) and the geometrical relation for 
$l$, the dimensionless height satisfies:

 \begin{equation}
h^* =  (R_l/R)^{3/2} \mathrm{.}
\label{petith}
\end{equation}

For usual liquids the value of $R_i$ is of a few hundred of microns. It is therefore worth mentioning that this prediction is qualitatively different and even opposed to the one commonly used for which the vapour thickness is expected to decrease when the radius of the droplet decreases.

The previous relations rely on the assumption that the the temperature field between the drop and the substrate (the plane located at $z=0$) is a solution of Laplace's equation. This approximation is valid in the limit in which  the convective heat flux is negligible compared to the diffusive flux (small Peclet number limit) and it holds in our experiments. 
From the image solution of potential problems, the temperature field between the drop and the hot plate is the same than the one between two spheres at temperatures $T_0$ and $2 T_1 -T_0$ and separated by a distance $2h$. Laplace's equation for two spheres was solved one hundred years ago by Jeffery \cite{jeffery} thanks to bi-spherical coordinates. We have numerically recovered the expected dependence of the temperature gradient  in the limits of $h/R$ large and $h/R$ small. As given above by scaling arguments, they respectively read $\nabla T \sim \Delta T R/h^2$ and $\nabla T \sim \Delta T /h$.

Considering a water drop on a hot plate at $400^0$C the numerical estimate of the critical radius is 
$R_l  \simeq  19 \mu{\rm m}$. To our knowledge no quantitative experiment 
has been done at such small sizes.  
 We therefore decided to perform experiments on ultra-distilled water and ethanol drops to verify the scaling laws derived just above.
  The experimental set-up is depicted in Fig. \ref{fig_set-up}. 
  A cloud of sub-millimetric droplets is rained on a silicon substrate kept at a controlled temperature $T_1$. 
  The two upper blocks are necessary to reduce the number of drops falling in the field of the camera.
They are also useful for the thermal stability of the system.
  Experiments are performed under a laminar air flow protection in order to prevent dust contamination. 
A high-speed camera, at a frame rate of $2000$ frame per second, was used to record several evaporation take-off processes. An image analysis allowed us to measure the radius and elevation of droplets as a function of time. The typical error made on these measurements  is  $2 \mu m$.
On the top of Fig. \ref{fig_ilus}, we show six pictures taken during the take-off of a water droplet. On the bottom of Fig. \ref{fig_ilus} we plot the height of the water droplet deposited on a silicon substrate at $T=375 ^{\circ}C$ and room pressure. It first bounces and lands on the silicon substrate. It then enters in the lubrication regimes in
which both radius and vapour thickness decrease with time. At certain time the droplet takes-off from the substrate (see the movie M1 in the supplementary materials).

\begin{figure}[!h]
  \includegraphics[width=.45\textwidth]{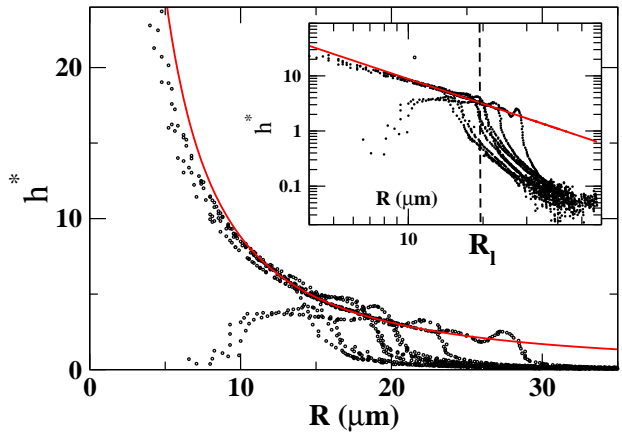}
  \caption{Dimensionless height of the droplets as a function of their radii.  The red line is a best fit to the equation (\ref{grandh}). Note that a minority of droplets fall back on the substrate. This is certainly due to dust contamination.
The same data are represented in the insert but in log-log. The dotted line indicates the value of $R_l$ for water Leidenfrost droplets on a substrate kept at $T=400 ^{\circ}C$.}
  \label{fig_eau}
\end{figure}

 We recorded several evaporation processes and we  represent in Fig. \ref{fig_eau} the dimensionless elevation of the droplet $h^*$ as a function of its radius $R$. 
The measurements are in qualitative agreement with the pictures of Fig. 2 (top) : almost all droplets take-off from the substrate when their radii becomes of order $R_l$. The theoretical prediction given above is tested against experiments.
The full line is a best fit with a single parameter $a$ multiplying the predicted critical radius $R_l$. We can see that the theoretical prediction fits rather well the experimental data with a value $a=2.2$ close to unity. One can see that there is an important dispersion in sizes at which the droplets take-off. It seems that, for the system of water droplets under study, the lift-off is analogous to a sub-critical transition (in other terms for the same radius and temperature difference, there are two equilibrium elevation in a certain range of values of $R/R_l$). The same behaviour has been observed for water droplets on substrates at different temperatures. Decreasing the temperature tends to decrease the dispersion
 of radius's at take-off. Nevertheless the discontinuous character of the escape from the lubrication regime remains.

\begin{figure}[!h]
  \includegraphics[width=.45\textwidth]{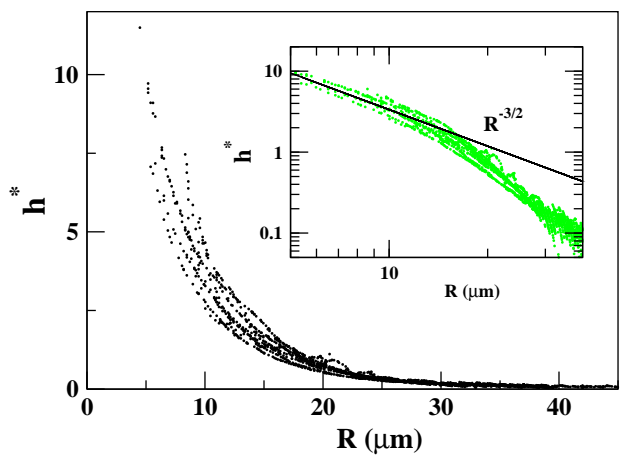}
  \caption{Dimensionless height of ethanol droplets as a function of $R$. The substrate is kept at $T=400 ^{\circ}C$. The same data are represented in the insert but in  log-log. The black  line  corresponds to the scaling law  predicted in  equations (\ref{grandh}). }
  \label{fig_ethanol}
\end{figure}

We  investigated also the behaviour of ethanol Leidenfrost droplet. The same experimental procedure is used and the silicon substrate is kept at
 similar temperatures than for the study of water Leidenfrost droplets. 
 Contrarily to water droplets almost all ethanol droplets do not rebound and land on the substrate (see the movie M2 in the supplementary materials). This is due to the lower latent heat of ethanol and therefore to its larger evaporation 
 rate. As a consequence the ethanol droplets enter directly in the regime described above for which $h>>l$ and we do not observe a transition from the lubrication
 regime. The experimental results are presented in Fig. \ref{fig_ethanol} in which we plot, as for water droplets, the dimensionless elevation of the droplet $h^*$ as a function of its radius $R$. The data are presented for $10$ different ethanol droplets on a substrate kept at $T=400 ^{\circ}C$. As stressed just above the escape from the lubrication regime is no longer present and the height $h$ of the droplets increases in a continuous manner. In the  insert of Fig. \ref{fig_ethanol} we represent the same experimental data but plotted in a log-log representation. The black  line corresponds to the scaling law predicted in  equation (\ref{grandh}). A rather good agreement is found between the theoretical prediction and the experiments.
 
 In this letter we have put in evidence the unexpected behaviour of a
 Leidenfrost droplet in the latter stage of its evaporation. Below a critical size $R_l$ the droplets
  escape from the lubrication regime and take-off from the substrate. In an intermediary regime 
  ($R_l \ll R \ll R_i$) the thickness of the vapour film increases as the radius of the droplet decreases.
This latter prediction, based upon a detailed analysis of the various
physical phenomena involved, is qualitatively different, and even opposed to the one
commonly used: as a result of the standard approach to the Leidenfrost
phenomenon the vapour thickness is predicted to {\emph{decrease}} as the radius of the droplet decreases.
 The different scaling laws presented
in this paper are in  semi-quantitative agreement with the experimental data
obtained on water and ethanol Leidenfrost droplets. In particular, as observed
and predicted, the thickness of the vapour film {\emph{increases}} as
the droplet become smaller. Besides
its fundamental interest this study should find implications in many
 domains such as for example in diesel combustion engines or in heat transfer using
 cooling spray \cite{kim}. We hope that this study will motivate
numerical approaches complementary to the present study.


\begin{thebibliography}{0}
\expandafter\ifx\csname natexlab\endcsname\relax\def\natexlab#1{#1}\fi
\expandafter\ifx\csname bibnamefont\endcsname\relax
  \def\bibnamefont#1{#1}\fi
\expandafter\ifx\csname bibfnamefont\endcsname\relax
  \def\bibfnamefont#1{#1}\fi
\expandafter\ifx\csname citenamefont\endcsname\relax
  \def\citenamefont#1{#1}\fi
\expandafter\ifx\csname url\endcsname\relax
  \def\url#1{\texttt{#1}}\fi
\expandafter\ifx\csname urlprefix\endcsname\relax\def\urlprefix{URL }\fi
\providecommand{\bibinfo}[2]{#2}
\providecommand{\eprint}[2][]{\url{#2}}

\end{thebibliography}


\begin{references}

\bibitem{leiden} J. G. Leidenfrost, {\em De Aquae Communis Nonnullis Qualitatibus Tractatus} (Duisbourg, 1756)

\bibitem{yves} L. Mahadevan and Y. Pomeau, Phys. Fluid {\bf 11}, 2449 (1999).

\bibitem{quere} A. L. Biance, C. Clanet and D. Qu\'er\'e, Phys. Fluid {\bf 15}, 1632 (2003) and references therein.

\bibitem{linke} H. Linke, B. J. Alem·n, L. D. Melling, M. J. Taormina, M. J. Francis, C. C. Dow-Hygelund, V. Narayanan, R. P. Taylor, and A. Stout, Phys. Rev. Lett.   {\bf 96}, 154502 (2006).

\bibitem{toro} I. U. Vakarelski, J. O. Marston, D. Y. C. Chan and S. T. Thoroddsen, Phys. Rev. Lett.   {\bf 106}, 214501 (2011).

\bibitem{lagubeau}  G. Lagubeau, M. le Merrer, C. Clanet and D. Qu\'er\'e, Nature Physics {\bf 7}, 395 (2011).

\bibitem{eggers} J. H. Snoeijer, P. Brunet and J. Eggers, Phys. Rev. E  
{\bf 79}, 036307 (2009).

\bibitem{gold} T. R. Cousins, R. E. Goldstein, J. W. Jaworski and A. I. Pesci.
J. Fluid. Mech. {\bf 79}, in press (2012).
(2009).

\bibitem{wurger} A. Wurger, Phys. Rev. Lett.   {\bf 107}, 164502 (2011).

\bibitem{note} 
We consider a film  below a  spherical drop of thickness $h(r)=h_0 + r^2/2R$.   
Here $r$ is the distance to the center of the drop in cylindrical coordinates and $h_0$ the film thickness at $r=0$.
Applying the standard lubrication approximation to describe the flow under the drop in evaporation leads to the following relation between $p(r)$ and $h(r)$ : 
\begin{equation}
\frac{d}{ dr} (\frac{rh^3}{12}\frac{dp}{dr}) + \frac{r \eta \lambda \delta T}{L \rho_v h} = 0.
\end{equation}
 After a double integration the  pressure field reads : 
\begin{equation}
p(r) = p_0- \frac{12\eta \lambda \delta T}{L \rho_v} \int_0^r \frac{dr_1}{r_1 h^3(r_1)} \int_0^{r_1} \frac{r_2 dr_2}{h(r_2)}
\end{equation} 
where $p_0$ is a constant so that the pressure decreases to zero at infinity.
Using the expression given just above for $h(r)$ it can be shown that the pressure decreases rapidly to $0$ as $r^{-6}$ on a scale length $\sqrt{Rh_0}$. Therefore $\sqrt{R h_0}$ is a fair
estimate of the horizontal extent of the layer where the viscous pressure due to the evaporation contributes significantly to the lifting force.

\bibitem{nagel} J. C. Burton, A.L. Sharpe, R. C. A. van der Veen, A. Franco and S. R. Nagel, arXiv:1202.2157v1 [cond-mat.soft] (2012).

\bibitem{celestini} F. Celestini and G. Kirstetter, Soft Matter {\bf 8}, 5992 (2012).

\bibitem{jeffery} G. B. Jeffery, Proc. Roy. Soc. London {\bf 87}, 109 (1912).

\bibitem{kim} J. Kim, Int. J. Heat and Fluid Flow {\bf 28}, 753767 (2007).



\end{references}
\end{document}